\documentclass[aps,prd,superscriptaddress,12pt,showpacs,notitlepage]{revtex4}
\usepackage{graphics}
\usepackage{graphicx}
\usepackage{amsfonts}
\usepackage{amsmath}

\newcommand{\be}{\begin{equation}}
\newcommand{\ee}{\end{equation}}
\newcommand{\bea}{\begin{eqnarray}}
\newcommand{\eea}{\end{eqnarray}}
\newcommand{\pa}{\partial}
\newcommand{\bb}{\bibitem}
 
\begin{document}
\title{Supersymmetric K field theories and defect structures}
\author{C. Adam}
\affiliation{Departamento de F\'isica de Part\'iculas, Universidad de Santiago de Compostela and Instituto Galego de F\'isica de Altas Enerxias (IGFAE) E-15782 Santiago de Compostela, Spain}
\author{J.M. Queiruga}
\affiliation{Departamento de F\'isica de Part\'iculas, Universidad de Santiago de Compostela and Instituto Galego de F\'isica de Altas Enerxias (IGFAE) E-15782 Santiago de Compostela, Spain}
\author{J. Sanchez-Guillen}
\affiliation{Departamento de F\'isica de Part\'iculas, Universidad de Santiago de Compostela and Instituto Galego de F\'isica de Altas Enerxias (IGFAE) E-15782 Santiago de Compostela, Spain}
\author{A. Wereszczynski}
\affiliation{Institute of Physics,  Jagiellonian University,
       \\ Reymonta 4, Krak\'{o}w, Poland}


\begin{abstract}
We construct supersymmetric K field theories (i.e., theories with a non-standard kinetic term) in 1+1 and 2+1 dimensions such that the bosonic sector just consists of a nonstandard kinetic term plus a potential. Further, we study the possibility of topological defect formation in these supersymmetric models.  Finally, we consider more general supersymmetric K field theories where, again, topological defects exist in some cases.
\end{abstract}

\maketitle 

\section{Introduction}
Topological defects are of fundamental importance in a wide range of physical theories. Both in particle theory and in condensed matter physics, topological defects may exist as stable, particle-like excitations above the ground state of a theory. In some cases,  states containing topological defects are even energetically preferred over the homogeneous state, such that the true ground state of the system is a condensate or lattice of topological defects. Another field where topological defects are deemed relevant is cosmology. On the one hand, 
topological defects are crucial in inflationary scenarios, where they may form domain walls separating different vacua of some primordial fields in the symmetry-breaking phase. As a consequence,
it is widely believed that a pattern of these topological defects might be responsible for the structure formation in  the very early universe, see e.g. \cite{Vil}, \cite{Hind}, \cite{Battye}. 
On the other hand, topological defects also play an important role in the so-called brane-world scenario, where it is assumed that the visible universe is a 3+1 dimensional subspace ("brane") in a higher-dimensional bulk universe. The brane may be either strictly 3+1 dimensional ("thin brane") or have a small but nonzero extension also in the additional dimensions ("thick brane"). In the latter, thick brane case, these branes are normally  topological defects in the higher-dimensional bulk space \cite{Akama1} - \cite{Dzhu1}. In all these cosmological applications, the relevant topological defects are usually solutions of some effective field theories of one or several scalar fields. The scalar field theories may either consist of the standard kinetic term of the scalar fields plus a potential, in which case the specific properties of the defects are related  to the properties of the potential. Or one may relax the condition on the kinetic term and allow for more general field theories with a Lagrangian depending both on the fields and their first derivatives. These so-called K field theories 
have been increasing in importance during the last years, beginning with the observation about a decade ago that they might be relevant for the solution of some problems in cosmology, like K-inflation
\cite{k-infl} and K-essence \cite{k-ess}. K field theories have found their applications in cosmology \cite{bab-muk-1} - \cite{trodden}, and they introduce some qualitatively new phenomena, like the formation of solitons with compact support, so-called compactons \cite{werle} - \cite{fring}.

If K field theories turn out to be relevant for the cosmological problems described above at sufficiently early times (e.g., in the inflationary epoch) and/or sufficiently small scales, then the question of supersymmetric extensions of these theories naturally arises (see, e.g., \cite{Rocher}, \cite{Yama}). Here the situation is quite different for theories supporting topological defects with standard kinetic term (possibly coupled to gauge fields), on the one hand, and K field theories, on the other hand. 
Standard scalar field theories (for co-dimension one defects), the abelian Higgs (or Chern--Simons Higgs) models (for co-dimensions two defects), the t´Hooft-Polyakov monopole  theory (for co-dimension 3 defects) and pure Yang--Mills theory (for co-dimension 4 defects) are all well-known to allow for supersymmetric extensions \cite{divecchia-ferrara}, \cite{witten-olive}, \cite{edelstein-nunez}, \cite{d'adda-horsley}, \cite{d'adda-divecchia}, and these supersymmetric extensions have been studied intensively over the last decades.

On the contrary, much less is known about supersymmetric extensions of K field theories supporting topological solitons.
To the best of our knowledge, the problem of supersymmetric extensions was first investigated in relation to the Skyrme model \cite{skyrme}, which is one of the best-known theories supporting topological solitons and possessing a non-standard kinetic term.
Concretely, the supersymmetric extensions of a $S^2$ (or CP(1)) restriction of the Skyrme model (the so-called Skyrme--Faddeev--Niemi (SFN) model \cite{FN1}) were investigated in \cite{nepo} and in \cite{frey}. In both papers, a formulation of the SFN model was used where the CP(1) restriction of the Skyrme model is achieved via a gauging of the third, unwanted degree of freedom. As a result, the SFN model is expressed by two complex scalar fields and an undynamical gauge field, which are then promoted to two chiral superfields and a real vector superfield in the Wess--Zumino gauge, respectively. 
The result of the analysis is that the SFN model in its original form cannot be supersymmetrically extended by these methods. Instead, the supersymmetric extension contains further terms already in the bosonic sector, and also the field equations of the bosonic fields are different.
Recently, we were able to show, using methods similar to the ones employed in the present article, that the baby Skyrme model in 2+1 dimensions does allow for a supersymmetric extension \cite{bS}. 
Quite recently, the investigation of the problem of possible supersymmetric extensions of scalar K field theories has gained momentum,  \cite{bazeia2}, \cite{susy1}, \cite{ovrut1}, \cite{ovrut2}. Here, \cite{bazeia2} and \cite{susy1} studied supersymmetric extensions of K field theories in 1+1 and in 2+1 dimensions, whereas the investigations of \cite{ovrut1} and \cite{ovrut2} are for 3+1 dimensional K theories, and related to some concrete cosmological applications (ghost condensates and Galileons).      

It is the purpose of the present article to introduce and study a large class of supersymmetric extensions of scalar K field theories as well as their static topological defect solutions. The supersymmetric field theories we construct exist both in 1+1 and in 2+1 dimensional Minkowski space, due to the similarity of the spin structure in these two spaces. The topological defect solutions we study, on the other hand, all will belong to the class of defects in 1+1 dimensions (kinks), or to co-dimension one defects in a more general setting. Concretely, in Section 2 we introduce a set of supersymmetric Lagrangians which we shall use as "building blocks" for the specific supersymmetric Lagrangians we want to construct. We find that it is possible to construct supersymmetric Lagrangians such that their bosonic sectors just consist of a generalized kinetic term plus a potential term. We also investigate stability issues (energy positivity and the null energy condition). In Section 3, we investigate topological defect solutions of the  theories introduced in Section 2. We find that there exist two classes of solutions, namely the so-called "generic" ones, which exist for a whole class of Lagrangians, and "specific" ones which depend on the specific Lagrangian under consideration. As these non-linear theories are rather uncommon, we discuss one prototypical example of the "specific" solutions in some detail. In this example, it results that all specific topological kink solutions belong to the class ${\cal C}^1$ of continuous functions with a continuous first derivative.  We then briefly discuss some further examples, where both compact solitons and ${\cal C}^\infty$ functions may be found among the specific solutions. In Section 4, we introduce and study a more general class of supersymmetric Lagrangians, where the bosonic sector no longer can be expressed as a sum of a generalized kinetic term and a potential. We also comment on the relation of our results with the results of Bazeia, Menezes, and Petrov \cite{bazeia2}. Finally, Section 5 contains a discussion of our results.

\section{Supersymmetric models}

\subsection{Conventions}

Our supersymmetry
conventions are based on the widely used ones of \cite{Siegel}, where
our only difference with their conventions is our choice of the
Minkowski space metric $\eta_{\mu\nu} = {\rm diag} (+,-,-)$ (or its restriction to 1+1 dimensions, where appropriate). All sign differences between this paper and \cite{Siegel} can be traced back to this difference.
Concretely, we use the superfield
\begin{equation}
\Phi (x,\theta ) =\phi (x) +\theta^\gamma\psi_\gamma (x) -\theta^2 F(x) 
\end{equation}
where $\phi$ is a real scalar field, $\psi_\alpha $ is a fermionic two-component Majorana spinor, and $F$ is the auxiliary field. Further, $\theta^\alpha$ are the two Grassmann-valued superspace coordinates,
and $\theta^2 \equiv (1/2)\theta^\alpha \theta_\alpha$. Spinor indices are risen and lowered with the spinor metric $C_{\alpha \beta} = -C^{\alpha\beta}=(\sigma_2)_{\alpha\beta}$, i.e., $\psi^\alpha = C^{\alpha\beta}\psi_\beta$ 
and $\psi_\alpha = \psi^\beta C_{\beta\alpha}$. 
The superderivative is
\begin{equation}
D_\alpha=\partial_\alpha+i\theta^\beta\partial_{\alpha\beta}=\partial_\alpha -i \gamma^\mu{}_\alpha{}^\beta \theta_\beta \partial_\mu
\end{equation}
and obeys the following useful relations ($D^2 \equiv \frac{1}{2} 
D^\alpha D_\alpha$):
\begin{eqnarray}
D_{\alpha}D_{\beta}=i\partial_{\alpha\beta}+C_{\alpha\beta}D^2\quad; 
\quad D^\beta D_\alpha D_\beta=0 \quad; \quad (D^2)^2=-\square \\
D^2 D_\alpha=-D_\alpha D^2=i\partial_{\alpha\beta}D^\beta \quad;\quad 
\partial^{\alpha\gamma}\partial_{\beta\gamma}=-\delta_\beta^\gamma\square
\end{eqnarray}
and
\begin{eqnarray}
D^2&=&\frac{1}{2}\partial^\alpha\partial_\alpha-i\theta^\alpha
\partial_{\beta\alpha}\partial^\beta+\theta^2\square \\
D^2\Phi&=&F-i\theta^\gamma\partial_{\delta\gamma}\psi^\delta+\theta^2
\square\phi\\
D^2(\Phi_1 \Phi_2)&=&(D^2\Phi_1)\Phi_2+(D^\alpha\Phi_1)(D_\alpha\Phi_2)+
\Phi_1D^2\Phi_2
\end{eqnarray}

The components of superfields can be extracted with the help of the following projections
\be
\label{comp}
\phi(x)=\Phi(z)|,\quad\,\psi_{\alpha}(x)=D_{\alpha}\Phi(z)|,\quad\,F(x)=D^2\Phi(z)|,
\ee
where the vertical line $|$ denotes evaluation at $\theta^\alpha =0$.

\subsection{Lagrangians}

For the models we will construct we need the following superfields 
\be
D^\alpha \Phi D_\alpha \Phi  = 2\psi^2 -2\theta^\alpha (\psi_\alpha F + i  \psi^\beta \pa_{\alpha\beta} \phi )
 + 2\theta^2 (F^2 -i\psi^\alpha \pa_{\alpha\beta} \psi^\beta + \pa_\mu \phi \pa^\mu \phi )
\ee
\bea
D^\beta D^\alpha \Phi D_\beta D_\alpha \Phi &=& 2\left( \pa_\mu \phi \pa^\mu \phi -\theta^\gamma \pa^{\alpha\beta} \psi_\gamma \pa_{\alpha\beta} \phi + \theta^2 \pa^{\alpha\beta} \phi \pa_{\alpha\beta}F + \frac{1}{2} \theta^2 \pa^{\alpha\beta} \psi^\gamma
\pa_{\alpha\beta} \psi_\gamma \right. \nonumber \\
&& \left. + F^2 -2i F \theta^\gamma \pa_{\delta\gamma} \psi^\delta +2\theta^2 F\Box \phi + \theta^2 \pa_\delta{}^\gamma \psi^\delta \pa_{\beta\gamma}\psi^\beta \right) 
\eea
\be
D^2 \Phi D^2 \Phi =  F^2 -2i F \theta^\gamma \pa_{\delta\gamma} \psi^\delta +2\theta^2 F\Box \phi + \theta^2 \pa_\delta{}^\gamma \psi^\delta \pa_{\beta\gamma}\psi^\beta 
\ee
as well as their purely bosonic parts (we remark that all spinorial contributions to the lagrangians we shall consider are at least quadratic in the spinors, therefore it is consistent to study the subsector with $\psi_\alpha =0$)
\bea
(D^\alpha \Phi D_\alpha \Phi)_{\psi =0} &=& 2 \theta^2 (F^2 + \partial^\mu \phi \partial_\mu \phi )  \label{D-eq-1}\\
(D^\beta D^\alpha \Phi D_\beta D_\alpha \Phi)_{\psi =0} &=& 2(F^2 + \partial^\mu \phi \partial_\mu \phi )  +
 4\theta^2 (F \Box \phi  - \pa_\mu \phi \pa^\mu F  )  
\label{D-eq-2} \\
(D^2\Phi D^2 \Phi)_{\psi =0} &=& F^2 +2 \theta^2 F \Box \phi  . \label{D-eq-3}
\eea
Next, let us construct the supersymmetric actions we want to investigate. A supersymmetric action always is the superspace integral of a superfield. Further, due to the Grassmann integration rules $\int d^2\theta  =0$, $\int d^2\theta \theta_\alpha =0$, $\int d^2\theta \theta^2 =-1=D^2 \theta^2$, the corresponding Lagrangian in ordinary space-time always is the $\theta^2$ component of the superfield. 
Besides, we are mainly interested in the bosonic sectors of the resulting theories, therefore we shall restrict to the purely bosonic sector in the sequel. We will use the following supersymmetric Lagrangian densities (in ordinary space-time) as building blocks,
\bea
( {\cal L}^{(k,n)})_{\psi =0} &=& -\left( D^2 [(\frac{1}{2} D^\alpha \Phi D_\alpha \Phi) (\frac{1}{2} D^\beta D^\alpha \Phi
D_\beta D_\alpha \Phi )^{k-1}(D^2 \Phi D^2 \Phi )^n ]| \right)_{\psi =0} \nonumber \\
&=& (F^2 +\pa_\mu \phi \pa^\mu \phi )^k F^{2n} \label{building-b}
\eea
where $k=1,2,\ldots$ and $n=0,1,2,\ldots$.
The idea now is to choose certain linear combinations of the ${\cal L}^{(k,n)}$ with specific properties. We observe that a general linear combination contains terms where powers of the auxiliary field $F$ couple to the kinetic term $\pa_\mu \phi \pa^\mu \phi$. But there exists a specific linear combination where these mixed terms are absent, namely
\begin{eqnarray}
({\cal L}^{(k)})_{\psi =0} &\equiv & 
({\cal L}^{(k,0)})_{\psi =0}-\binom{k}{1}({\cal L}^{(k-1,1)})_{\psi =0}+\binom{k}{2}({\cal L}^{(k-2,2)})_{\psi =0}+ \ldots\\ \nonumber
&& \ldots +(-1)^{k-1}\binom{k}{k-1}({\cal L}^{(1,k-1)})_{\psi =0}= (\partial^\mu\phi\partial_\mu\phi)^k + (-1)^{k-1}F^{2k} . 
\label{k-n-sum}
\end{eqnarray}
The Lagrangians we want to consider are linear combinations of the above, where we also want to include a potential term, because we are mainly interested in topological solitons and defect solutions. That is to say, we add a prepotential $P(\Phi)$ to the action density in superspace which, in ordinary space-time, induces the bosonic Lagrangian density $(D^2 P)| = P'(\phi) F$ (here the prime denotes a derivative w.r.t. the argument $\phi$). Therefore, the class of Lagrangians we want to consider is
\bea
{\cal L}_b^{(\alpha ,P)} &=& \sum_{k=1}^N \alpha_k ({\cal L}^{(k)})_{\psi =0} + P'  F \nonumber \\
&=& \sum_{k=1}^N \alpha_k [ (\partial^\mu\phi\partial_\mu\phi)^k + (-1)^{k-1}F^{2k}] + P'(\phi) F
\eea
where the lower index $b$ means "bosonic" (we only consider the bosonic sector of a supersymmetric Lagrangian), and the upper index $\alpha$ should be understood as a multiindex $\alpha = (\alpha_1 ,\alpha_2 , \ldots , \alpha_N )$ of coupling constants. Further, $N$ is a positive integer. In a next step, we eliminate $F$ via its algebraic field equation
\be \label{F-eq}
\sum_{k=1}^N (-1)^{k-1} 2k \alpha_k F^{2k-1} + P' (\phi )=0.
\ee
For a given function $P(\phi)$ this is, in general, a rather complicated algebraic equation for $F$. However, we made no assumption yet about the functional dependence of $P$, therefore we may understand this equation in a second, equivalent way: we assume that $F$ is an arbitrary given function of $\phi$, which in turn determines the prepotential $P(\phi)$.  This second way of interpreting Eq. (\ref{F-eq}) is more useful for our purposes. Eliminating the resulting $P'(\phi)$ we arrive at the Lagrangian density
\bea \label{L-b-F}
{\cal L}_b^{(\alpha ,F)}
&=& \sum_{k=1}^N \alpha_k [ (\partial^\mu\phi\partial_\mu\phi)^k - (-1)^{k-1}(2k-1) F^{2k}] 
\eea
where now $F=F(\phi)$ is a given function of $\phi$ which we may choose freely depending on the theory or physical problem under consideration. 

\subsection{Energy considerations}

We would like to end this section with some considerations on the positivity of the energy. The energy density corresponding to the Lagrangian 
(\ref{L-b-F}) is (in 1+1 dimensions and with $\dot \phi = \pa_t \phi$, $\phi ' = \pa_x \phi$)
\be
{\cal E}_b^{(\alpha ,F)} = \sum_{k=1}^N \alpha_k \left( (\dot \phi^2 - \phi '^2)^{k-1} ((2k-1)\dot\phi^2 + \phi '^2) + (-1)^{k-1} (2k-1) F^{2k} \right)  .
\ee
This expression is obviously positive semi-definite if only the $\alpha_k$ with odd $k$ are nonzero and positive. It remains positive semi-definite if both the lowest (usually $k=1$) and the highest value of $k$ ($k=N$) with a nonzero and positive $\alpha_k$ are odd, provided that the intermediate $\alpha_k$ for even $k$ are not too large. For a given value of $N$, inequalities for the coefficients $\alpha_k$ guaranteeing positive semi-definiteness of the energy density can be derived without difficulties.

A second, less restrictive condition which is deemed sufficient to guarantee stability is the so-called null energy condition
\be
T_{\mu\nu} n^\mu n^\nu \ge 0
\ee
where $T_{\mu\nu}$ is the energy-momentum tensor and $n^\mu$ is an arbitrary null vector. For general Lagrangians ${\cal L}(X,\phi)$ where $X\equiv \frac{1}{2} \pa_\mu \phi \pa^\mu \phi$, the energy momentum tensor reads
\be
T_{\mu\nu} = {\cal L}_{,X} \pa_\mu \phi \pa_\nu \phi - g_{\mu\nu} {\cal L}
\ee
(here ${\cal L}_{,X}$ is the $X$ derivative of ${\cal L}$), and the null energy condition simply is ${\cal L}_{,X} \ge 0$. 
For our specific class of Lagrangians (\ref{L-b-F}), the null energy condition therefore reads
\be
({\cal L}_b^{(\alpha ,F)})_{,X} = \sum_{k=1}^N 2k\alpha_k (\pa_\mu \phi \pa^\mu \phi )^{k-1} \ge 0.
\ee
Again, this condition is automatically satisfied if only $\alpha_k$ for odd $k$ are nonzero, or if the $\alpha_k$ for even $k$ obey certain restrictions. 

Remark: in \cite{susy1} a class of models based on the superfield (\ref{chi}) of Section 4.2 were introduced. These models satisfy neither energy positivity nor the null energy condition. They support, nevertheless, topological kink solutions, and their energy densities can be expressed as the squares of the corresponding supercharges.  A more complete analysis of these models which would resolve the issue of stability is, therefore, an open problem at the moment which requires further investigation. 
    
\section{Solutions}

The Euler-Lagrange equation for the Lagrangian density (\ref{L-b-F}) is 
\be
\sum_{k=1}^N 2k \alpha_k \left( \pa_\mu [ (\pa_\nu \phi \pa^\nu \phi )^{k-1} \pa^\mu \phi ] + (-1)^{k-1} (2k-1) F^{2k-1} F_{,\phi} \right) =0
\ee
which, in 1+1 dimensions and for static (time-independent) fields simplifies to
\be \label{gen-eq}
\sum_{k=1}^N 2k (-1)^{k-1} \alpha_k \left( -\pa_x (\phi_{,x}^{2k-1} ) + (2k-1) F^{2k-1} F_{,\phi} \right) =0 .
\ee

\subsection{Generic static solutions}
First of all, we want to demonstrate that, due to the restrictions imposed by supersymmetry, this equation has a class of static, one-dimensional solutions which are completely independent of the coefficients $\alpha_k$, which we shall call the "generic" solutions. Indeed, if we impose equation (\ref{gen-eq}) for each $k$ (i.e., for each term in the sum) independently, the resulting equation is
\be
\pa_x (\phi_{,x}^{2k-1}) = (2k-1) F^{2k-1}F_{,\phi} 
\ee
or, after multiplying by $\phi_{,x}$ and dividing by $(2k-1)$,
\be
\phi_{,x}^{2k-1} \phi_{,xx} = F^{2k-1}F_{,\phi} \phi_{,x}
\ee
which may be integrated to $\phi_{,x}^{2k}= F^{2k}$ and, therefore, to the $k$ independent solution
\be \label{indep-F-eq}
\phi_{,x} \equiv \phi ' =\pm F.
\ee
That is to say, these solutions only depend on the choice of $F=F(\phi)$, but do not depend on the $\alpha_k$ and, therefore, exist 
for an infinite number of theories defined by different values of the $\alpha_k$. Depending on the choice for $F(\phi)$, the static solutions may be topological solitons. E.g. for the simple choice $F=1-\phi^2$, the solution of  (\ref{indep-F-eq}) is just the well-known $\phi^4$ kink solution
$\phi (x) = \tanh (x-x_0)$ where $x_0$ is an integration constant (the position of the kink). As another example, for $F=\sqrt{|1-\phi^2|}$, we get the compacton solution 
\begin{equation}
\phi (x) = \left\{
\begin{array}{lc}
- 1 & x-x_0 \leq - \frac{\pi}{2} \\
\sin (x - x_0) & \quad
-\frac{\pi}{2} \leq x-x_0 \leq
\frac{\pi}{2}  \\
1 & x-x_0 \geq \frac{\pi}{2}
\end{array}
\right. \label{compacton-sol}
\end{equation}
where, again, $x_0$ is an integration constant.

The energy of a generic supersymmetric kink solution may be calculated with the help of the first order formalism, which has the advantage that an explicit knowledge of the kink solution is not needed for the determination of its energy (for details on the first order formalism we refer to \cite{bazeia3}).
All that is needed is the field equation of a generic solution $\phi ' = \pm F$ (we shall choose the plus sign corresponding to the kink, for concreteness). The idea now is to separate a factor $\phi '$ in the energy density with the help of the  field equation, because this allows to rewrite the base space integral of the energy functional as a target space integral with the help of the relation $\phi ' dx \equiv d\phi$. Concretely, we get for the energy density of the generic kink solution
\bea
{\cal E} &=& \sum_{k=1}^N (-1)^{k-1} \alpha_k (\phi '^{2k} + (2k-1) F^{2k}) = \sum_{k=1}^N (-1)^{k-1} \alpha_k 2k F^{2k} 
\nonumber \\
&=& 
\phi ' \sum_{k=1}^N (-1)^{k-1} \alpha_k 2k F^{2k-1} \equiv \phi ' W_{,\phi}
\eea
where $W_{,\phi}$ and its $\phi $ integral $W (\phi)$ are understood as functions of $\phi$. For the energy this leads to
\be
E = \int_{-\infty}^\infty dx \phi ' W_{,\phi} = \int_{\phi (-\infty)}^{\phi (\infty )} d\phi W_{,\phi} = W(\phi (\infty)) - W(\phi (-\infty)).
\ee
 As indicated, all that is needed for the evaluation of this energy is the root $\phi ' = F(\phi)$ and the asymptotic behaviour $\phi (\pm\infty)$ of the kink. We remark that the integrating function $W(\phi)$ of the first order formalism is identical to the prepotential $P(\phi)$, 
\be
W(\phi)=P(\phi)
\ee
as is obvious from Eq. (\ref{F-eq}). This is exactly as in the case of the standard supersymmetric scalar field theory with the standard, quadratic kinetic term. It also remains true for the class of models introduced and studied in \cite{bazeia2}, as we shall discuss in some more detail in Section 4.B. Both for the standard supersymmetric scalar field theories and for the models introduced in \cite{bazeia2} it is, in fact, possible to rewrite the energy functional for static field configurations in a BPS form, such that both the first order field equations for static fields and the simple, topological expressions $E=P(\phi(\infty))-P(\phi(-\infty))$ for the resulting energies are a consequence of the BPS property of the energy functional (for the models introduced in \cite{bazeia2} we briefly recapitulate the BPS property of static kink solutions in Section 4.B). On the contrary, for the models introduced in the previous section there is no obvious way to rewrite them in a BPS form,
despite the applicability of the first order formalism, because the energy functional contains, in general, many more than two terms (just two terms are needed to complete a square and arrive at the BPS form). 

On the other hand, for the additional, specific solutions of the theories of Section 2 to be discussed in the following two subsections, the relation $W=P$  is no longer true, although it is still possible to calculate the energies of the specific solutions with the help of the first order formalism.

\subsection{Specific solutions: an example}
Next, we want to study whether in addition to the solutions $\phi_{,x}^2 = F^2$, which do not depend on the specific Lagrangian (i.e., on the coefficients $\alpha_k$), there exist further (static) solutions which do depend on the Lagrangian.  
Both the existence of such additional solutions and their properties (e.g., being topological solitons) will depend on the Lagrangian, therefore the results will be less general and have to be discussed separately for each model. So,
 let us select a specific Lagrangian (specific values for the $\alpha_k$) as an example. Concretely, we want to study the simplest case which gives rise to a potential with several vacua and obeys certain additional restrictions (positivity of the energy). Positivity of the energy requires that both the highest and the lowest nonzero $\alpha_k$ are for odd $k$, so we choose nonzero $\alpha_3$ and $\alpha_1$ for the simplest case. Further, we want that the potential factorizes and gives rise to several vacua, so we choose the concrete example $\alpha_3 = \frac{1}{5}$, $\alpha_2 = \frac{2}{3}$, and $\alpha_1 = 1$, which gives rise to the Lagrangian density
\be \label{alpha-lag}
{\cal L}_b^{ex} = \frac{1}{5} (\pa_\mu \phi \pa^\mu \phi )^3 + \frac{2}{3}  (\pa_\mu \phi \pa^\mu \phi )^2 + \pa_\mu \phi \pa^\mu \phi  
- F^6 + 2 F^4 - F^2 
\ee
where, indeed, the potential in terms of $F$ factorizes, $F^6 - 2 F^4 + F^2 = F^2 (1-F^2)^2$.   Next, we want to assume the simplest relation between $F$ and $\phi$, namely $F^2 (\phi )=\phi^2$. The resulting Lagrangian is
\be 
{\cal L}_b^{ex} = \frac{1}{5} (\pa_\mu \phi \pa^\mu \phi )^3 + \frac{2}{3}  (\pa_\mu \phi \pa^\mu \phi )^2 + \pa_\mu \phi \pa^\mu \phi  
- \phi^2 (1-\phi^2)^2 . 
\ee
We already know that it gives rise to the static solutions 
\be \label{exp-sol}
(\phi_{,x})^2 = \phi^2 \Rightarrow \phi (x) = \exp \pm (x-x_0).
\ee 
These solutions have infinite energy and are not solitons. We want to investigate whether there exist additonal solutions and, specifically, whether there exist topological solitons. The potential has the three vacua $\phi = (0,1,-1)$, therefore topological solitons (static solutions which interpolate between these vacua) are not excluded. 
We shall find that these solitons exist in the space ${\cal C}^1$ of continuous functions with a continuous first derivative, but not in the spaces ${\cal C}^n$ (with continuous first $n$ derivatives) for $n>1$.

\begin{figure}[h!]
\includegraphics[angle=0,width=0.55 \textwidth]{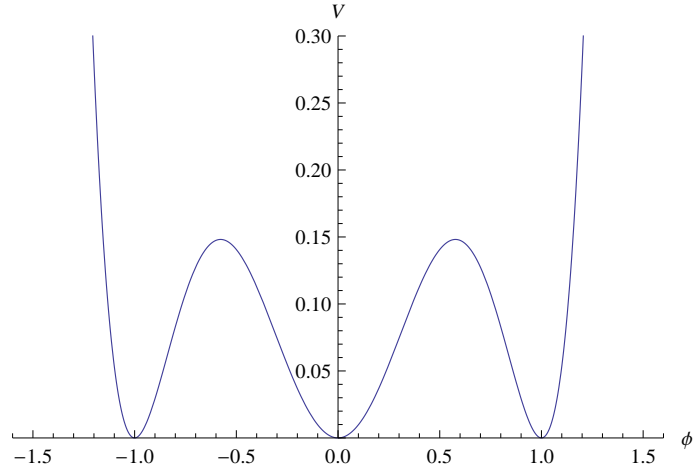}
\caption{The potential $V(\phi)= \phi^6 - 2 \phi^4 + \phi^2$.  }\label{fig-pot}
\end{figure}

The once-integrated field equation for static solutions (with the integration constant set equal to zero, as required by the finiteness of the energy) reads ($\phi ' \equiv \phi_{,x}$)
\be \label{example-eq}
\phi '^6 -2\phi '^4 + \phi '^2 = \phi^6 -2 \phi^4 + \phi^2 
\ee
and obviously has the solutions (\ref{exp-sol}). For a better understanding of further solutions the following observations are useful. Firstly, for a fixed value $x=\tilde x$ of the independent variable $x$, the field $\phi$ and its derivative $\phi '$ have to obey the equation
\be
V(\phi) = V(\phi ')=c
\ee
where $c$ is a real, positive constant (or zero) and 
\be
V(\lambda)\equiv \lambda^6 - 2 \lambda^4 + \lambda^2 = \lambda^2 (1-\lambda^2)^2
\ee
is the potential (see Figure (\ref{fig-pot})). In general, the equation $V(\lambda)=c$ has six solutions $\lambda = \pm \lambda_i (c), i = 1 \ldots 3$.  In other words, if we choose the initial condition $\phi (\tilde x) =\tilde \phi$, then $\phi '(\tilde x)$ is not uniquely determined (as would be the case for a linear first order equation) and may take any of the six values $\pm \lambda_i (c)$ such that $V(\pm\lambda_i (c)) = V(\tilde \phi) =c$. Obviously, the choice $\phi '(\tilde x) = \pm \tilde \phi$ leads to the exponential solutions (\ref{exp-sol}), whereas other choices will lead to additional solutions.

Secondly, the field equation (\ref{example-eq}) leads to the following equation for the second derivative 
\be
\phi '' = \frac{(3\phi^4 - 4 \phi^2 +1)\phi}{3\phi '^4 - 4 \phi '^2 +1}
\ee
where the numerator is zero for all critical points (minima and maxima) $\phi = (0,\pm 1, \pm \frac{1}{\sqrt{3}})$ of the potential $V(\phi)$, whereas the denominator is zero for the critical points $\phi ' = (\pm 1, \pm \frac{1}{\sqrt{3}})$. For later convenience we also remark that at the two local maxima $\phi = \pm \frac{1}{\sqrt{3}}$ the potential takes the value $V(\pm \frac{1}{\sqrt{3}}) = \frac{4}{27}$ and that the equation
$V(\lambda)=\frac{4}{27}$ has the two further solutions $\lambda = \pm \frac{2}{\sqrt{3}}$ which are not critical points.

We observe that the equation $V(\phi ')= V(\phi)$ can in fact be solved algebraically for $\phi '$ and leads to the solutions 
\be \label{phi-prime-eq0}
\phi ' = \pm \phi
\ee
(that is, the exponential solutions (\ref{exp-sol})) and to the four further solutions
\be \label{phi-prime-eq}
\phi ' = \pm \frac{1}{2} \left( \phi \pm \sqrt{ 4-3\phi^2}\right) .
\ee
For this last expression, reality of $\phi '$ requires that $|\phi | \le \frac{2}{\sqrt{3}}$. The resulting integral for $\phi$
\be \label{imp-sol}
\int_0^\phi \frac{d\tilde \phi}{\pm ( \phi \pm \sqrt{4-3\phi^2})} = \frac{1}{2}(x-x_0)
\ee
may be resolved explicitly, providing an implicit solution $x-x_0 = H(\phi)$ where, for each choice of signs, $H(\phi)$ is a combination of logaritms  and  inverse trigonometric functions. The explicit expressions for $H$ are, however, rather lenghty and not particularly illuminating, therefore we prefer to continue our discussion with a combination of qualitative arguments and numerical calculations. We want to remark, however, that the graphs of the numerical solutions shown in the figures below agree exactly with the graphs of the analytic solutions (\ref{imp-sol}) (we remind the reader that for the graph of a function the implicit solution is sufficient).

For the qualitative discussion, we now assume that we choose an "initial value" at a given point $\tilde x$. Due to translational invariance we may choose this point at zero $\tilde x =0$, i.e. $\phi (0)= \phi_0$, without loss of generality. For $0<|\phi_0| < \frac{1}{\sqrt{3}}$,  $\frac{1}{\sqrt{3}} < |\phi_0| < 1$ and $1< |\phi_0| < \frac{2}{\sqrt{3}}$, $\phi ' (0)$ may take any of the six real solutions of the equation $V(\phi ') = V(\phi_0)$. Further, $\phi ''(0)$ as well as all higher order derivatives at $x=0$ are uniquely determined by linear equations, as we shall see in a moment. Therefore, for these "initial conditions" $\phi_0$, there exist indeed the six solutions (\ref{exp-sol}) and (\ref{imp-sol}). 
For $|\phi_0|>\frac{2}{\sqrt{3}}$, only the two solutions $\phi'(0)= \pm \phi_0$ of the equation $V(\phi ') = V(\phi_0)$ are real, therefore only the two exponential solutions (\ref{exp-sol}) exist. At the critical points $\phi_0 =0,\pm1$ and $\phi_0 = \pm \frac{1}{\sqrt{3}},\pm \frac{2}{\sqrt{3}}$ the situation is slightly more complicated (strictly speaking $\phi_0 =\pm \frac{2}{\sqrt{3}}$ are not critical points, because $V'(\pm \frac{2}{\sqrt{3}}) \not= 0$; however, $\phi_0 =\pm \frac{2}{\sqrt{3}}$ provides the same level height of the potential like the critical points
$\phi_0 = \pm \frac{1}{\sqrt{3}}$, that is, $V( \pm \frac{1}{\sqrt{3}}) = V( \pm \frac{2}{\sqrt{3}}) = \frac{4}{27}$, therefore these points play a special role in the analysis, too). In order to understand what happens it is useful to insert the Taylor expansion about $x=0$,
\be
\phi (x) = \sum_{k=0}^\infty f_k x^k
\ee
into the field equation $V(\phi) - V(\phi ')=0$, which, up to second order, reads
\bea
0 &=& f_0^2 (1-f_0^2)^2 - f_1^2 (1-f_1^2)^2 + \\
&& [2f_0 (1-4f_0^2 + 3 f_0^4 )f_1 -4 f_1 (1-4f_1^2 + 3f_1^4)f_2 ] x + \\
&& [(1-12f_0^2 + 15 f_0^4 )f_1^2 + 2f_0 (1-4f_0^2 + 3 f_0^4 )f_2 - \nonumber \\
&& - 4(1-12 f_1^2 + 15 f_1^4 )f_2^2 - 6f_1 (1-4f_1^2 + 3 f_1^4 )f_3 ] x^2 + \ldots 
\eea
It can be inferred easily that for generic values of $f_0$ and $f_1$ (values which are not critical points), $f_2$ is determined uniquely by a linear equation from the term of order $x^1$. On the other hand, if $f_1$ takes a critical value, then the coefficient multiplying $f_2$ in the order $x^1$ term is zero, and $f_2$ is determined, instead, by a quadratic equation coming from the term of order $x^2$. These points will be important in the following, because precisely at these points we may join different solutions such that the resulting solution belongs to the class ${\cal C}^1$ of continuous functions with a continuous first derivative.
Specifically, we find the following possible values for $f_1$ and $f_2$ for a given, critical $f_0$ (we only consider the cases $f_0\ge 0$ because of the obvious symmetry $\phi \to -\phi$ of the theory). For $f_0 =0$
\be
(f_0,f_1,f_2) \; = \; (0,0,0) \; \; \mbox{or} \; \; (0,\pm 1 ,\pm \frac{1}{4}) 
\ee
where the first case corresponds to the trivial vacuum solution $\phi \equiv 0$, and the second case corresponds to the four solutions (\ref{imp-sol}). The exponential solutions (\ref{exp-sol}) are obviously incompatible with the "initial condition" $\phi (0)=0$ (the vacuum solution $\phi (x) =0$ can be understood as a limiting case of the two exponential solutions for infinite integration constant $x_0$).  Next, for $f_0 =1$
\be
(f_0,f_1,f_2) \; = \;  (1,0,0)  \; \;\mbox{or} \; \; (1,\pm 1 ,\pm \frac{1}{2}) 
\ee
where the first case corresponds to the trivial vacuum solution $\phi \equiv 1$. The second case consists of the exponential solutions (two solutions) and of two of the four solutions (\ref{imp-sol}). The other two are incompatible with the "initial conditions". For $f_0 = \frac{1}{\sqrt{3}}$ we get 
\be
(f_0,f_1,f_2) \; = \;  (\frac{1}{\sqrt{3}},\pm \frac{1}{\sqrt{3}},\pm \frac{1}{2\sqrt{3}}) \; \; \mbox{or} \; \; (\frac{1}{\sqrt{3}},\pm \frac{2}{\sqrt{3}} ,0) 
\ee
where the first case contains both the two exponential solutions and two of the four solutions (\ref{imp-sol}), and the second case corresponds to the other two solutions (\ref{imp-sol}). Finally, for $f_0 = \frac{2}{\sqrt{3}}$ we find
\be
(f_0,f_1,f_2) \; = \;  (\frac{2}{\sqrt{3}},\pm \frac{2}{\sqrt{3}},\frac{1}{\sqrt{3}}) \; \; \mbox{or} \; \; (\frac{2}{\sqrt{3}},\pm \frac{1}{\sqrt{3}}, \infty )
\ee
where the first case provides the two exponentail solutions (\ref{exp-sol}), whereas the second case shows that the solutions (\ref{imp-sol}) run into a singularity when $|f_0 | =\frac{2}{\sqrt{3}}$. 

\begin{figure}[h!]
\includegraphics[angle=0,width=0.45 \textwidth]{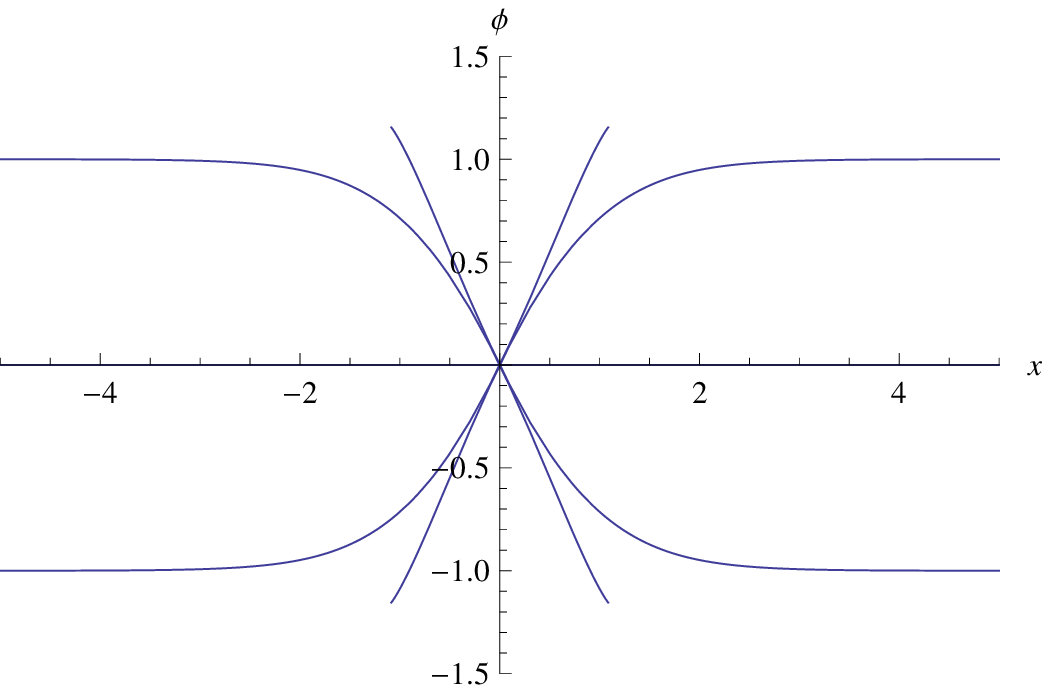}
\includegraphics[angle=0,width=0.45 \textwidth]{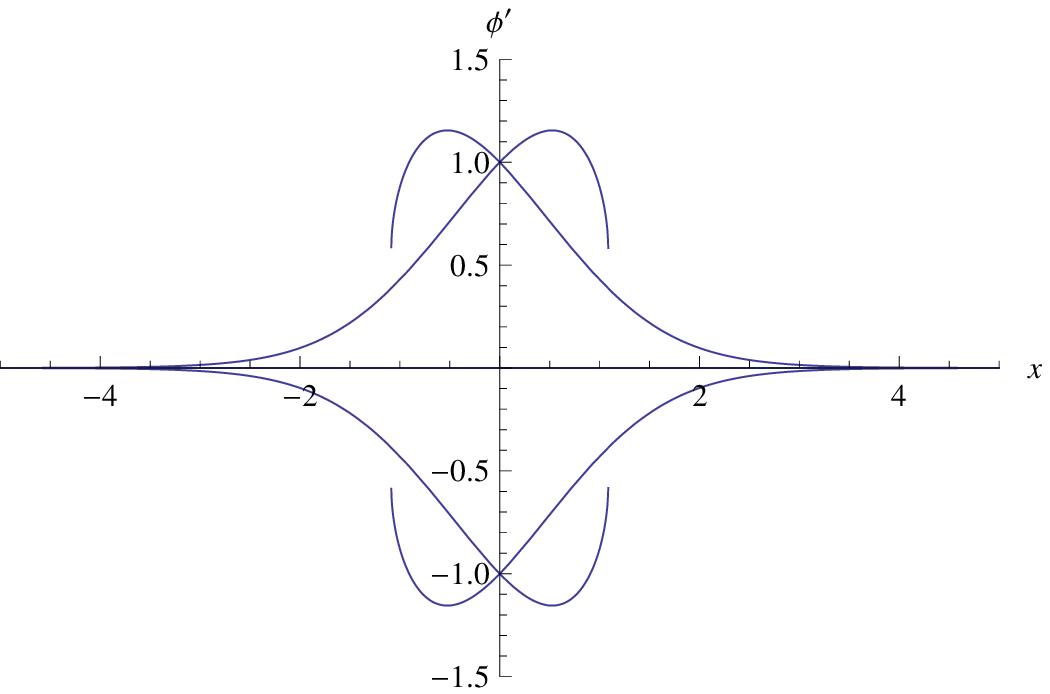}
\caption{For the "initial condition" $\phi (0)=0$ all the five solutions (including the trivial solution $\phi \equiv 0$) $\phi (x)$ (left figure) and the first derivatives $\phi '(x)$ (right figure). The singularity at $\phi (x_2) =\pm \frac{2}{\sqrt{3}}$, $\phi '(x_2)= \pm\frac{1}{\sqrt{3}}$ for some $x_2$, where the integration breaks down, is clearly visible.   }\label{phi01}
\end{figure}

Now let us study some of these cases in more detail. Concretely, we investigate the case $(f_0,f_1,f_2) =  (0, 1 , \frac{1}{4})$.  Firstly, for negative $x$, $\phi(x)$ diminishes from $\phi (0)=0$ towards $-1$, and $\phi '(x)$ diminishes from $\phi '(0)=1$ towards $0$, such that for a fixed value of $x$ $\phi$ and $\phi '$ have the same height on the graph of $V$, see Fig. (\ref{fig-pot}).  If $x$ is sufficiently negative such that $\phi (x)$ is close to its vacuum value $-1$ and $\phi ' (x)$ is close to zero, the field equation may be linearized about the vacuum $-1$, and it follows easily that the vacuum is approached exponentially, like $\phi (x) \sim -1+ \exp (4x)$ (remember that $x$ is negative). In other words, for negative $x$ the solution behaves like a nice kink or topological soliton and does not reach the vacuum value $-1$ for finite $x$. For positive $x$, in a first instant both $\phi (x)$ and $\phi '(x)$ grow till they reach the values $\phi (x_1)=\frac{1}{\sqrt{3}}$ and $\phi '(x_1) = \frac{2}{\sqrt{3}}$ for some $x_1>0$. At this point $\phi ''(x_1)=0$ therefore $\phi '$ may change direction in a smooth way. For $x>x_1$, $\phi$ continues to grow while $\phi '$ shrinks till they reach the values $\phi (x_2)=\frac{2}{\sqrt{3}}$ and $\phi '(x_2)=\frac{1}{\sqrt{3}}$ for some $x=x_2$. At this point $\phi ''(x_2)=\infty$, and the solution hits a singularity. A numerical integration confirms these findings, see Figure (\ref{phi01}).

\begin{figure}[h!]
\includegraphics[angle=0,width=0.45 \textwidth]{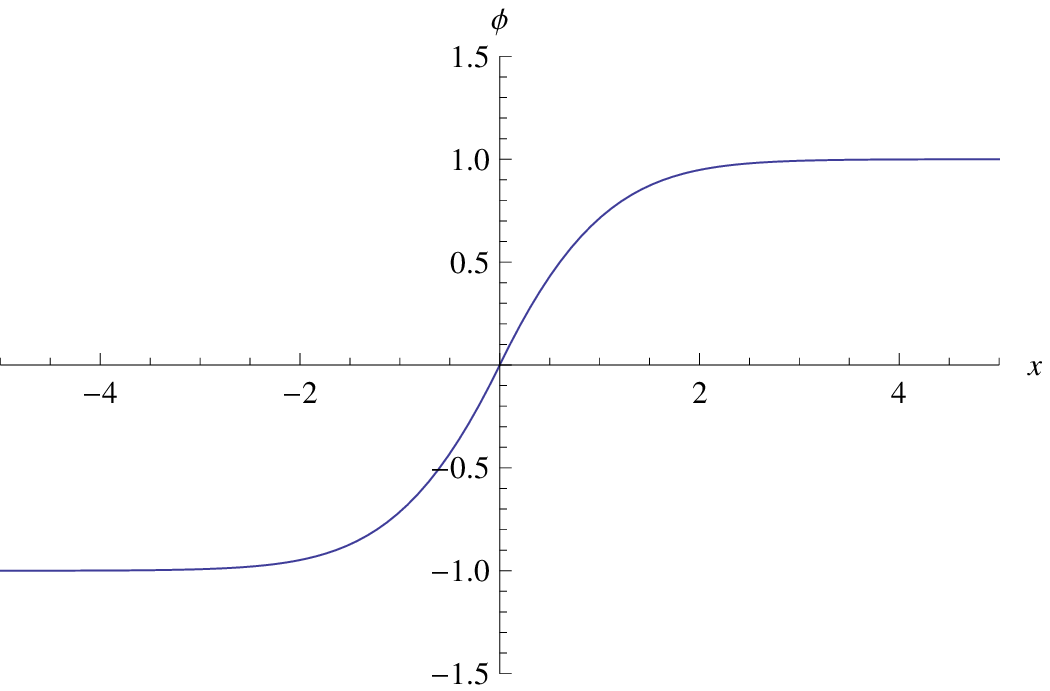}
\includegraphics[angle=0,width=0.45 \textwidth]{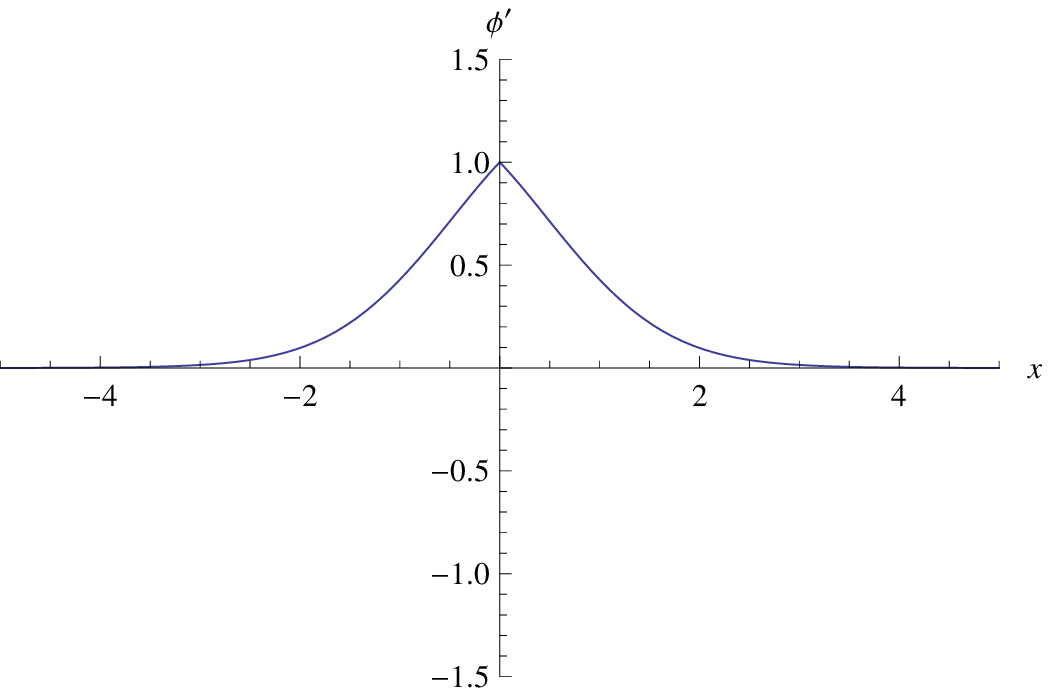}
\caption{For the "initial condition" $\phi (0)=0$, the kink solution interpolating between $\phi =-1$ and $\phi =1$ (left figure)  and its first derivative (right figure).  }\label{phi01kink}
\end{figure}

There exists, however, the possibility to form a topological soliton or kink solution in the class ${\cal C}^1$ of continuous functions with continuous first derivatives by simply joining the solution $(f_0,f_1,f_2) =  (0, 1 , \frac{1}{4})$  for negative $x$ with the solution $(f_0,f_1,f_2) =  (0, 1 , -\frac{1}{4})$ for positive $x$. Indeed, both $\phi (0)$ and $\phi '(0)$ agree, so the resulting solution is ${\cal C}^1$. Further, $\phi '$ in the second case diminishes for positive $x$ because $\phi ''(0)$ is negative. Therefore, $\phi (x)$ approaches $1$ and $\phi '$ approaches $0$ for large positive $x$, and a linearized analysis reveals that in that region $\phi (x) \sim 1- \exp (-4x)$. As a consequence, the solution obtained by the joining procedure behaves exactly like a kink interpolating between the vacuum  $-1$ at $x=-\infty$ and the vacuum $+1$ at $x=\infty$. For the corresponding result of a numerical integration, see Figure (\ref{phi01kink}).

\begin{figure}[h!]
\includegraphics[angle=0,width=0.45 \textwidth]{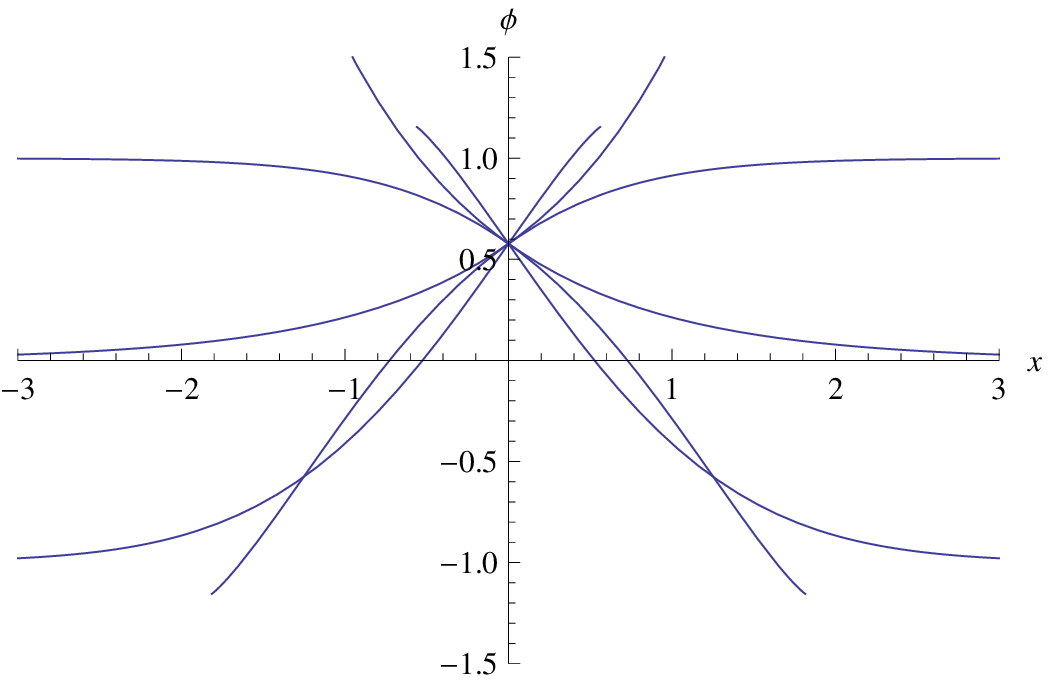}
\includegraphics[angle=0,width=0.45 \textwidth]{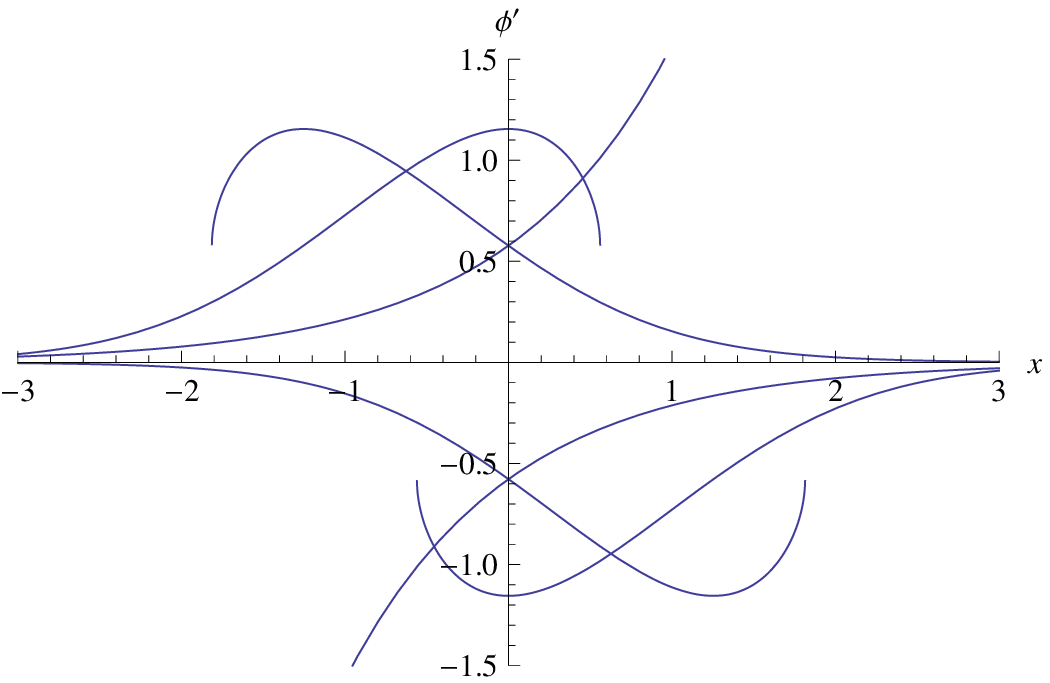}
\caption{For the "initial condition" $\phi (0)=\frac{1}{\sqrt{3}}$ all six solutions $\phi (x)$ (left figure) and the first derivatives $\phi '(x)$ (right figure). Again, the singularities at $\phi (x_2) =\pm \frac{2}{\sqrt{3}}$, $\phi '(x_2)= \pm\frac{1}{\sqrt{3}}$ for some $x_2$ for the non-exponential solutions are clearly visible.   }\label{phisq3}
\end{figure}

\begin{figure}[h!]
\includegraphics[angle=0,width=0.45 \textwidth]{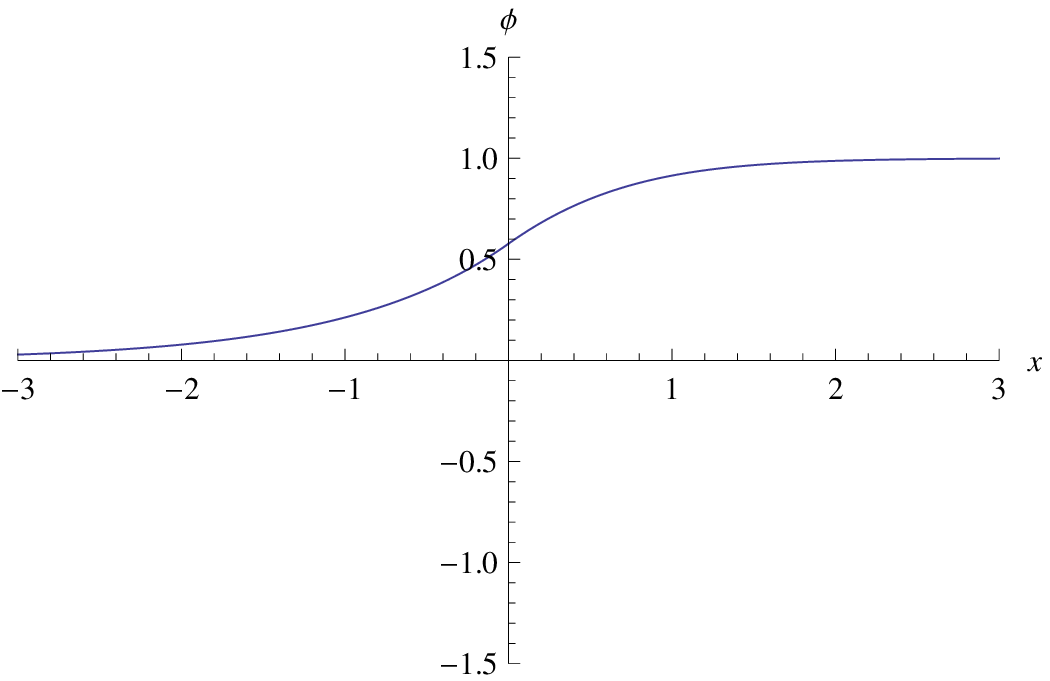}
\includegraphics[angle=0,width=0.45 \textwidth]{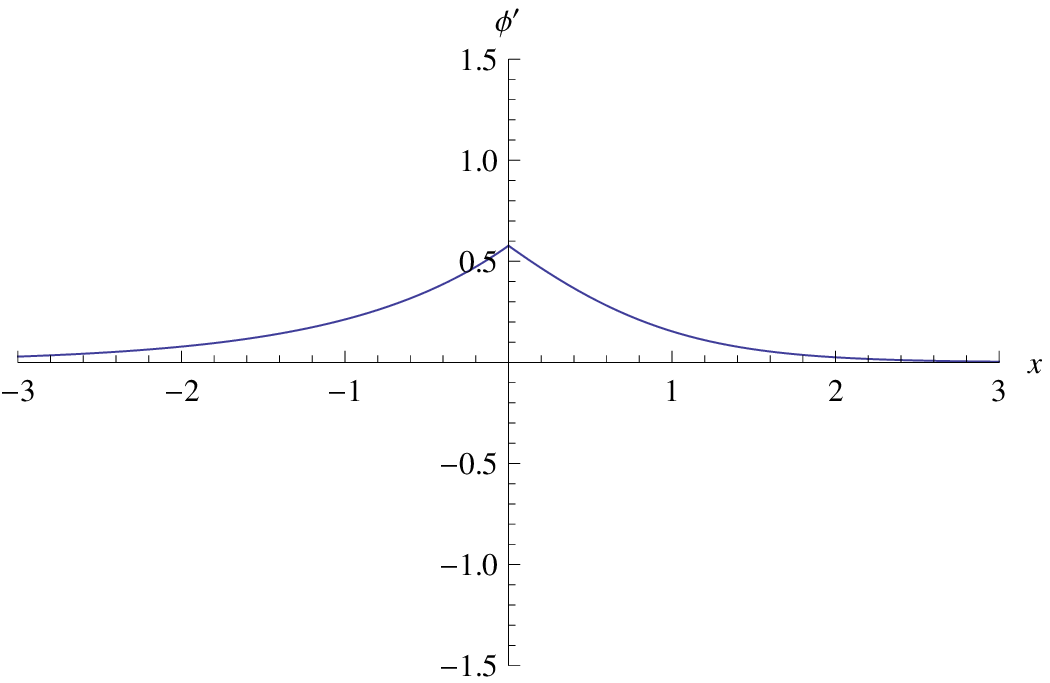}
\caption{For the "initial condition" $\phi (0)=\frac{1}{\sqrt{3}}$, the kink solution interpolating between $\phi =0$ and $\phi =1$ (left figure)  and its first derivative (right figure).  }\label{phisq3kink}
\end{figure}

Finally, let us discuss the possibility to form a kink in the class of ${\cal C}^1$ functions which interpolates, e.g., between the vacuum 0 and the vacuum 1. For this purpose, we should join the solution $(f_0,f_1,f_2)=(\frac{1}{\sqrt{3}},\frac{1}{\sqrt{3}},\frac{1}{2\sqrt{3}})$ for $x<0$ with the solution $(f_0,f_1,f_2)=(\frac{1}{\sqrt{3}},\frac{1}{\sqrt{3}},-\frac{1}{2\sqrt{3}})$ for $x>0$. Indeed, the solution 
$(f_0,f_1,f_2)=(\frac{1}{\sqrt{3}},\frac{1}{\sqrt{3}},\frac{1}{2\sqrt{3}})$ is just the exponential solution $\exp x$ and behaves well (approaches 0 exponentially) for negative $x$. For the solution $(f_0,f_1,f_2)=(\frac{1}{\sqrt{3}},\frac{1}{\sqrt{3}},-\frac{1}{2\sqrt{3}})$, on the other hand, both $\phi (0)$ and $\phi ' (0)$ are equal to $\frac{1}{\sqrt{3}}$ at $x=0$. For increasing $x$, $\phi (x)$ increases and $\phi '(x)$ decreases until they get close to 1 and 0, respectively. But near these values, again, a linearized analysis applies and tells us that $\phi$ behaves like
$\phi (x) \sim 1-\exp (-4x)$. Therefore, the solution produced by the joining procedure describes a kink which interpolates between the vacuum $\phi =0$ at $x=-\infty$   and the vacuum $\phi =1$ at $x=\infty$. The general solution for the initial condition $\phi (0)= \frac{1}{\sqrt{3}}$  is displayed in Figure (\ref{phisq3}), 
and the kink solution is shown in Figure (\ref{phisq3kink}).

The remaining kink and antikink solutions which we have not discussed explicitly may be easily found with the help of the obvious symmetries $x\to -x$ and $\phi \to -\phi $. We remark that from the point of view of the variational calculus, solutions in the ${\cal C}^1$ class of functions are perfectly valid. They provide well-defined energy densities and, therefore, well-defined critical points of the energy functional. Whether they are acceptable from a physics point of view depends, of course, on the concrete physical problem under consideration.

Finally, we want to calculate the energies of the kinks constructed by the joining procedure described above. These energies can again be calculated exactly and do not require the knowledge of the explicit solutions $\phi (x)$ but, instead, just the knowledge of the six roots 
(\ref{phi-prime-eq0}), (\ref{phi-prime-eq}) of the first order equation $V(\phi) = V(\phi ')$, Eq. (\ref{example-eq}). Indeed, with the help of the first order equation we find for the energy density for static solutions
\be
{\cal E} = -{\cal L}_b^{ex} = \frac{1}{5}\phi'^6 - \frac{2}{3} \phi '^4 + \phi'^2 + \phi^2 (1-\phi^2)^2 =
\frac{6}{5}\phi'^6 - \frac{8}{3} \phi '^4 + 2 \phi '^2 
\ee
and for the energy
\bea 
E &=& \int_{-\infty}^\infty dx {\cal E} = \int_{-\infty}^\infty dx \phi ' ( \frac{6}{5}\phi'^5 - \frac{8}{3} \phi '^3 + 2 \phi ' ) \nonumber \\
&\equiv & \int_{-\infty}^\infty dx \phi ' W_{,\phi} = \int_{\phi (-\infty)}^{\phi (\infty)} d\phi W_{,\phi} = W(\phi (\infty)) - W(\phi (-\infty))
\eea
where 
\be \label{W}
W_{,\phi} (\phi) = \frac{6}{5}\phi'^5 - \frac{8}{3} \phi '^3 + 2 \phi '
\ee
(and its $\phi$ integral $W(\phi)$)
must be understood as a function of $\phi$ which results when evaluating the above expression for one of the six roots (\ref{phi-prime-eq0}), (\ref{phi-prime-eq}) for $\phi'$. Here we use again the first order formalism of \cite{bazeia3} to which refer for a detailed discussion. 

In our case, the kinks are constructed by joining two different solutions, therefore the expression for the energy is slightly more complicated and reads
\be
E = W^{(2)} (\phi (\infty)) - W^{(2)} (\phi (0)) + W^{(1)} (\phi (0)) - W^{(1)} (\phi (-\infty))
\ee
where $W^{(2)}$ and $W^{(1)}$ are the functions which result from evaluating the expression (\ref{W}) for the two different roots $\phi '$ which form the specific kink solution, and from performing the corresponding $\phi$ integrals. The joining may be done at any point $x_0$ in base space (because of translational invariance) but we chose $x_0 =0$ in our specific examples, therefore the joining point in target space is $\phi (0)$. 

Concretely, for the soliton of Figure 3 which interpolates between $\phi (-\infty)=-1$ and $\phi (\infty )=1$, with joining point $\phi (0)=0$, the correct roots are
\be
\phi <0 \, : \; \phi ' = \frac{1}{2} (\phi + \sqrt{4-3\phi^2}) \, , \qquad \phi > 0 \, : \; \phi ' = -\frac{1}{2}(\phi - \sqrt{4-3\phi ^2}).
\ee
Further, positive and negative $\phi$ regions give exactly the same contribution to the total energy, therefore the soliton energy can be calculated to be
\bea 
E_{(-1,1)} &=&  2 \int_{-1}^0 d\phi \left(  \frac{6}{5} [\frac{1}{2}(\phi + \sqrt{4-3\phi^2})]^5 - \frac{8}{3} [\frac{1}{2}(\phi + \sqrt{4-3\phi^2})]^3 + \phi + \sqrt{4-3\phi^2} \right) \nonumber \\
&=& 2 \left[ \frac{\phi^6}{10} - \frac{\phi^4}{12} + \left( \frac{\phi}{10} + \frac{11}{60} \phi^3 - \frac{\phi^5}{10} \right) \sqrt{4-3\phi^2} + \frac{2}{3\sqrt{3}} \arcsin \left( \frac{\sqrt{3}}{2}\phi \right) \right]_{\phi =-1}^0 \\
&=& 2\left( \frac{1}{6} + \frac{2\pi}{9\sqrt{3}} \right) = \frac{1}{3} + \frac{4\pi}{9\sqrt{3}}.
\eea
For the kinks which interpolate between $0$ and $\pm 1$, we choose the one which interpolates between $\phi (-\infty)=-1$ and $\phi (\infty)=0$, because then we may use exactly the same solution as above for the region between $\phi =-1$ and $\phi =-\frac{1}{\sqrt{3}}$. For $\phi$ between $  -\frac{1}{\sqrt{3}}$ and $0$, the correct root is $\phi ' =-\phi$, therefore we get for the total energy of this kink
\bea 
E_{(-1,0)} 
&=&  \left[ \frac{\phi^6}{10} - \frac{\phi^4}{12} + \left( \frac{\phi}{10} + \frac{11}{60} \phi^3 - \frac{\phi^5}{10} \right) \sqrt{4-3\phi^2} + \frac{2}{3\sqrt{3}} \arcsin \left( \frac{\sqrt{3}}{2}\phi \right) \right]_{\phi =-1}^{-\frac{1}{\sqrt{3}}} \nonumber  \\
&& + \left[ -\frac{1}{5}\phi^6 + \frac{2}{3} \phi ^4 - \phi^2 \right]_{\phi = -\frac{1}{\sqrt{3}}}^0 \nonumber \\
&=&  \frac{1}{6} + \frac{2\pi}{9\sqrt{3}} -\frac{7}{45} -\frac{\pi}{9\sqrt{3}} +\frac{4}{15} = \frac{5}{18} + \frac{\pi}{9\sqrt{3}}.
\eea
The remaining kinks, obviously, have the same energy. We remark that $E_{(-1,1)} >2 E_{(-1,0)}$. Therefore, the kink interpolating between $-1$ and $1$ probably is unstable against the decay into one kink interpolating between $-1$ and $0$ plus one kink interpolating between $0$ and $+1$. Establishing this conjecture would, however, require a numerical integration of the time-dependent system, which is beyond the scope of the present article.

\subsection{Further examples of specific solutions}

In this subsection we shall discuss two more examples which are similar to the theory studied in the last subsection.  
In the first example, the main difference is that the kinks no longer approach their vacuum values in an exponential fashion. Instead, two of the three vacua are approached compacton-like (i.e. the field takes the corresponding vacuum value already for finite $x$), whereas the third vacuum is approached in a power-like way (concretely like $\phi \sim x^{-1}$).  
In the second example, we will find that there exists a specific kink solution which belongs to the class of ${\cal C}^\infty$ functions. In both examples, we use the same values for the $\alpha_i$ like in (\ref{alpha-lag}). Besides, these examples are similar in many respects to the one discussed above in detail, so the discussion which follows can be much shorter. Also the resulting soliton energies can be calculated analytically, using exactly the same method like in the above example, therefore we do not repeat this calculation. 

In the first example, we choose for $F$
\begin{equation}
F=\sqrt{|1-\phi^2|} .
\end{equation}
We already know that this model leads to the $\alpha$-independent static field equations
\begin{equation} \label{comp-roots}
\phi' = \pm F = \pm \sqrt{|1-\phi^2|}
\end{equation}
which have the compacton solutions (\ref{compacton-sol}) (the corresponding anti-compacton solutions for the minus sign). The once integrated static field equation is
\begin{equation} \label{V-tilde-eq}
\phi'^{6}-2\phi'^4+\phi'^2=|1-\phi^2 |^3-2(1-\phi^2)^2+|1-\phi^2 | =\phi^4 |(1-\phi^2)|
\end{equation}
and might lead to further solutions, as in the previous subsection. Indeed, the potential $\tilde V = \phi^4  |1-\phi^2 |$ has the three vacua $\phi = 0,\pm 1$. Further, the potential $\tilde V$ behaves like $\tilde V \sim |\delta \phi |$ near the two vacua $\pm 1$ (i.e., for $\phi = \pm (1-\delta \phi )$ and small $\delta \phi$), whereas it behaves like $\tilde V \sim \delta \phi ^4$ near the vacuum $0$ (i.e. for $\phi = \delta \phi$ and small $\delta \phi$). The asymptotic field equations for $\delta \phi$ are $\delta \phi '^2 \sim |\delta \phi |$ near the two vacua $\pm 1$, with the asymptotic compacton-like solution  
$$\delta \phi \sim \frac{1}{2} (x-x_0)^2 \quad  \mbox{for} \quad x\le x_0 \; ; \quad \delta \phi =0 \quad  \mbox{for} \quad x>x_0.$$ 
The asymptotic field equation near the vacuum 0 is $\delta \phi'^2 \sim \delta \phi^4$ with the solutions  
\begin{equation} \nonumber
\delta \phi = \pm \frac{1}{x-x_0} +o\left(\frac{1}{x}\right), \;\;\;\; |x| \rightarrow \infty
\end{equation}
and, therefore, the algebraic, power-like localization announced above. It only remains to determine whether it is possible to join a solution with this  asymptotic behaviour to a compacton-like solution, forming a kink of the semi-compacton type, which interpolates, e.g., between the vacuum $\phi (-\infty) =0$ (with a power-like approach) and the vacuum $\phi (x_1)=1$ (where $x_1$ is the compacton boundary). For this purpose we note that Eq. (\ref{V-tilde-eq}) has, in addition to the two roots (\ref{comp-roots}), the four roots
\bea
\phi' &=& \pm \frac{1}{\sqrt{2}} \sqrt{1+\phi^2 + \sqrt{1+2\phi^2-3\phi^4} } \\
\phi' &=& \pm \frac{1}{\sqrt{2}} \sqrt{1+\phi^2 - \sqrt{1+2\phi^2-3\phi^4} }  . \label{neg-root}
\eea
The solution with the right behaviour (i.e., approaching the vacuum $\phi =0$) is the lower one, Eq. (\ref{neg-root}). Now we just have to determine whether it is possible to join this solution with the compacton solution such that both $\phi$ and $\phi '$ coincide at the joining point.
The result is that this joining is indeed possible, as may be checked easily. For the kink interpolating between $\phi (-\infty) =0$ and $\phi (x_1)=1$, e.g., the joining point is $\phi (x_0) = \sqrt{\frac{2}{3}}$, $\phi ' (x_0)= \frac{1}{\sqrt{3}}$ where, as always, the joining point  $x_0$ in base space is arbitrary. Further, for joining point $x_0$, the compacton boundary of the semi-compacton is at $x_1=x_0 + \frac{\pi}{2} - \arcsin \sqrt{\frac{2}{3}}$. 
 
We remark that in this example all kink solutions (both the compactons and the semi-compactons) are solutions in the ${\cal C}^1$ class of functions, because the second derivative of the field at the compacton boundary is not uniquely determined (its algebraic equation has three solutions, corresponding to the vacuum, the compacton, and a third solution with infinite energy, respectively). For the semi-compacton, the second derivative of the field at the joining point obviously is not uniquely determined, as well, analogously to the kinks formed by the joining procedure in the previous subsection.

\begin{figure}[h!]
\includegraphics[angle=0,width=0.45 \textwidth]{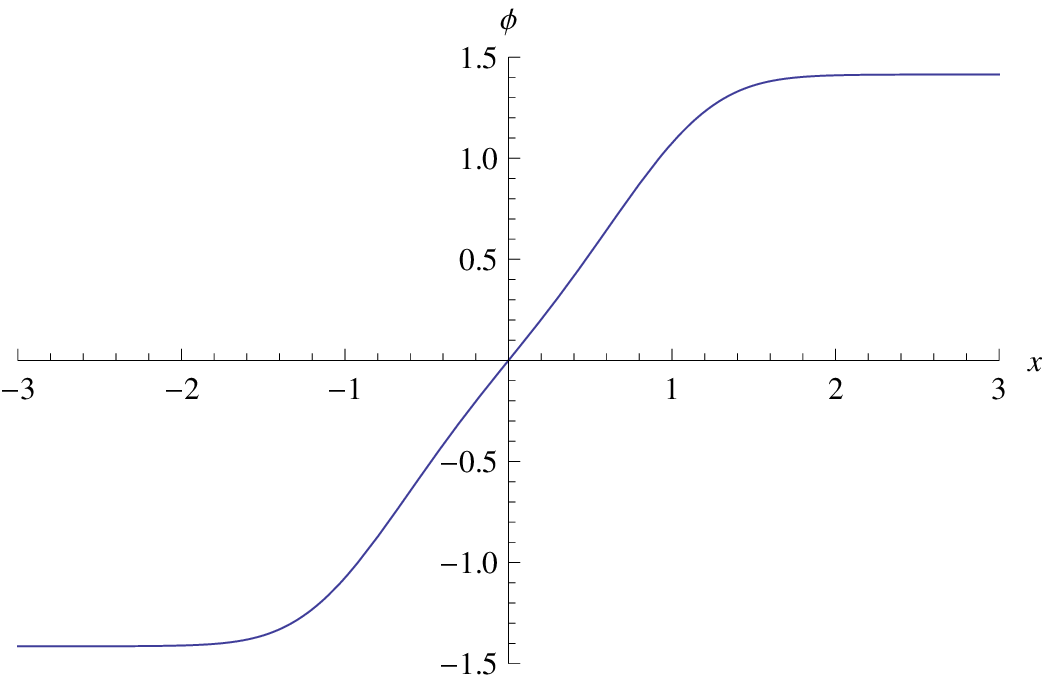}
\includegraphics[angle=0,width=0.45 \textwidth]{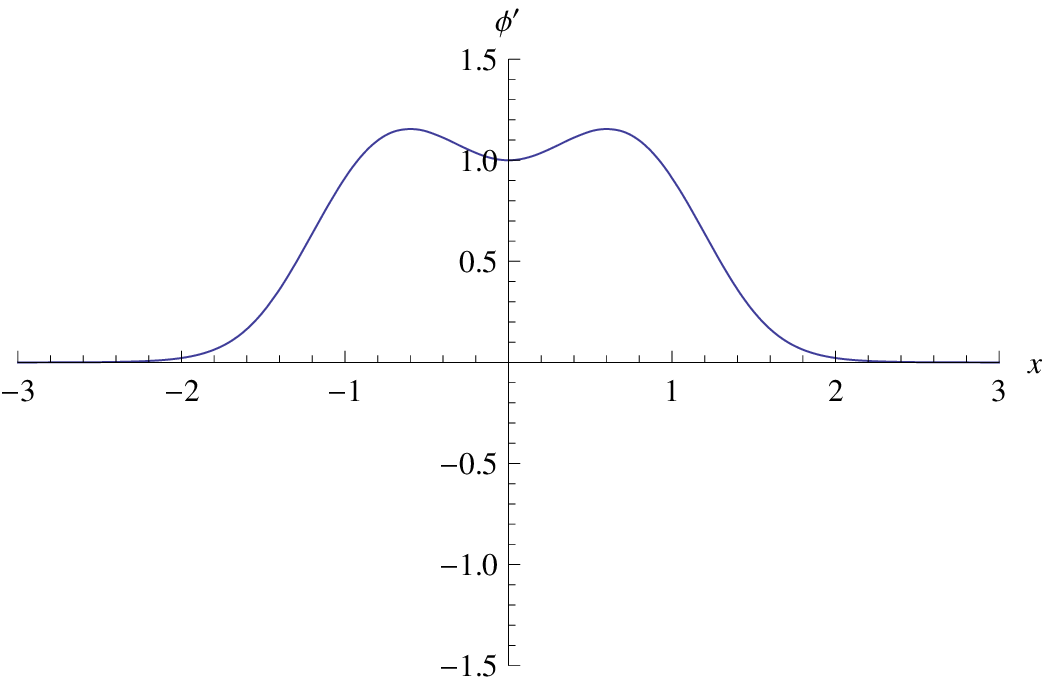}
\caption{The specific, regular kink solution interpolating between the two vacua $\phi = -\sqrt{2}$ and $\phi = \sqrt{2}$ (left figure) and its first derivative (right figure).  }\label{fig-kink22}
\end{figure}

For the second example we choose
\begin{equation}
F=1-\phi^2
\end{equation}
which leads to the following first order equation 
\begin{equation} \label{V-tildetilde-eq}
\phi'^{6}-2\phi'^4+\phi'^2=(1-\phi^2)^6-2(1-\phi^2)^4+(1-\phi^2)^2 =\phi^4(1-\phi^2)^2(2-\phi^2)^2 .
\end{equation}
Therefore, the resulting potential $\tilde{\tilde V} = \phi^4(1-\phi^2)^2(2-\phi^2)^2$ has the five vacua $\phi^2 = (0,1,2)$. Further, the four vacua $\phi = \pm 1$ and $\phi = \pm \sqrt{2}$ are approached quadratically and will, therefore, lead to the usual exponential kink tail. The vacuum $\phi =0$, on the other hand, is approached with a fourth power, $\tilde{\tilde V} \sim \delta \phi ^4$, and will lead to an algebraic, power-like tail, like in the last example. Concretely, we find near $\phi =0$, $\delta \phi '^2 \sim 4 \delta \phi^4$ and therefore
\begin{equation}
\delta \phi \sim \pm \frac{1}{2(x-x_0)}, \;\;\;\; |x| \rightarrow \infty .
\end{equation}
The six roots of Eq. (\ref{V-tildetilde-eq}) are 
\begin{equation}
\phi'=\pm ( 1-\phi^2)
\end{equation}
(which is just the first order equation for the standard $\phi^4$ kink) and 
the four additional equations 
\bea
\phi' &=& \pm \frac{1}{2} \left( 1-\phi^2 + \sqrt{1+6\phi^2-3\phi^4} \right) \label{2-2-kink} \\
\phi' &=& \pm \frac{1}{2} \left( 1-\phi^2 - \sqrt{1+6\phi^2-3\phi^4} \right) . \label{0-0-kink}
\eea
Here, equation (\ref{2-2-kink}) describes solutions which approach the two vacua $\phi = \pm \sqrt{2}$, whereas equation (\ref{0-0-kink})  describes solutions which approach the vacuum $\phi =0$. By joining different solutions, we may create kinks in the class ${\cal C}^1$ which interpolate between any two different vacua of the model, as we did in the previous two examples. Two of the kinks belong, however, to the class of ${\cal C}^\infty$ functions. The first ${\cal C}^\infty$ kink is just the standard $\phi^4$ kink interpolating between the two vacua $\phi =-1$ and $\phi =1$, and it respresents a generic solution of the model. The second ${\cal C}^\infty$ kink is the one interpolating between $\phi =-\sqrt{2}$ and $\phi = \sqrt{2}$, as we want to demonstrate now. Indeed, equation (\ref{2-2-kink}) describes a solution which approaches the two vacua $\phi = \pm \sqrt{2}$. Further, this equation (we choose the root with the plus sign, for concreteness) is completely regular in the interval $-\sqrt{2}\le \phi \le \sqrt{2}$ (the equation develops singularities at the two points $\phi^2 = 1 + \frac{2}{\sqrt{3}}$, but these points are outside the interval where the kink takes its values). Therefore, we expect that this equation should describe a smooth ${\cal C}^\infty$ kink which interpolates between the two vacua. Both an exact, implicit integration and a numerical integration precisely confirm this expecation. We display the result of a numerical integration in Fig. (\ref{fig-kink22}).

We remark that this kink has the interesting feature that its first derivative (and, therefore, also the energy density) has a local minimum at the position of the kink center, whereas the two local maxima are slightly displaced to the right and left of the center.  We further remark that this example demonstrated explicitly that ${\cal C}^\infty$ kinks may exist not only among the generic solutions but also among the specific solutions of our supersymmetric K field theories (which was not obvious in the other two examples studied so far).

\section{Further models}
\subsection{Field-dependent $\alpha_k$}
Here we want to construct further supersymmetric K field theories based on the observation that the models introduced in the last section remain supersymmetric when the factors $\alpha_k$ multiplying each power of the kinetic term depend on $\phi$ instead of being constants. Indeed, a superfield which is just an arbitrary function $\alpha (\Phi)$ of the basic superfield $\Phi$ has the superspace expansion in the bosonic sector $\psi =0$
\be
(\alpha (\Phi))_{\psi =0} = \alpha (\phi ) - \theta^2 \alpha ' (\phi) F
\ee
(the prime $\alpha '(\phi )$ denotes the derivative w.r.t. the argument $\phi$). If this superfield is multiplied by the superfield 
$(D^\alpha \Phi D_\alpha \Phi)_{\psi =0}$ which only has a $\theta^2$ component in the bosonic sector, then in the product only the $\theta$-independent component of $\alpha (\Phi)$ contributes,
\be
 (\alpha (\Phi))_{\psi =0}| = \alpha (\phi ) 
\ee
and the multiplication with the superfield $\alpha (\Phi)$ corresponds to a multiplication with the ordinary field $\alpha (\phi)$ of the Lagrangian densities (\ref{building-b}), i.e., to the new building blocks 
 \bea
( {\cal L}^{(k,n)})_{\psi =0} &=& -\left( D^2 [\alpha^{(k,n)}(\Phi) (\frac{1}{2} D^\alpha \Phi D_\alpha \Phi) (\frac{1}{2} D^\beta D^\alpha \Phi
D_\beta D_\alpha \Phi )^{k-1}(D^2 \Phi D^2 \Phi )^n ]| \right)_{\psi =0} \nonumber \\
&=& \alpha^{(k,n)}(\phi) (F^2 +\pa_\mu \phi \pa^\mu \phi )^k F^{2n} \label{building-b2}
\eea
where again $k=1,2,\ldots$ and $n=0,1,2,\ldots$. The cancellation of the mixed terms $ (\pa_\mu \phi \pa^\mu \phi )^i F^{2j}$ may again be achieved by calculating the sum analogous to (\ref{k-n-sum}) provided all the $\alpha^{(k,n)}$ in the sum are equal. Linear combinations of these Lagrangians are therefore 
\bea
{\cal L}_b^{(\alpha ,P)}
&=& \sum_{k=1}^N \alpha_k (\phi) [ (\partial^\mu\phi\partial_\mu\phi)^k + (-1)^{k-1}F^{2k}] + P'(\phi) F
\eea
just like in Section 2, but now with $\phi$ dependent coefficients $\alpha_k (\phi)$. Also, the field equation for the auxiliary field $F$ is the same and leads to the Lagrangian
\bea \label{L-b-F2}
{\cal L}_b^{(\alpha ,F)}
&=& \sum_{k=1}^N \alpha_k (\phi ) [ (\partial^\mu\phi\partial_\mu\phi)^k - (-1)^{k-1}(2k-1) F^{2k}] 
\eea
where $F=F(\phi)$ is an arbitrary function of $\phi$, like in (\ref{L-b-F}), but now with field dependent coefficients $\alpha_k(\phi)$. 
This result provides us with a new class of supersymmetric K field theories where now different powers of the kinetic term may be multiplied by functions of the scalar field.  

Now we shall discuss an explicit example, where we choose the functions $\alpha_k (\phi)$ and $F(\phi)$ such that the resulting model possesses a simple defect solution. Concretely we choose the nonzero $\alpha_k$
\be \label{alphas}
\alpha_3 = \frac{1}{24} \; , \quad \alpha_2 = \frac{1}{4}(1-\phi^2)^2 \; ,\quad \alpha_1 = \frac{1}{2}[1+ (1-\phi^2)^4] .
\ee
For non-constant $\alpha_k (\phi)$ positivity of the energy and the null energy condition become slightly more involved. For the moment we only consider the null energy condition, which is satisfied for this specific model. Indeed, the resulting Lagrangian is 
\be
{\cal L} = \frac{1}{24}[(\pa_\mu \phi \pa^\mu \phi )^3 - 5 F^6 ] + \frac{1}{4}(1-\phi^2)^2 [(\pa_\mu \phi \pa^\mu \phi )^2 +3F^4 ]
+ \frac{1}{2}[1+(1-\phi^2)^4](\pa_\mu \phi \pa^\mu \phi -F)
\ee
and for ${\cal L}_{X}$ we get
\be
{\cal L}_X = X^2 + 2 (1-\phi^2)^2 X + (1-\phi^2 )^4 +1 = (X+ (1-\phi^2)^2)^2 +1>0
\ee
(remember $X\equiv \frac{1}{2}\pa_\mu \phi \pa^\mu \phi$), so the null energy condition holds. In order to have simple defect solutions we now choose for $F$
\be
F^2= (1-\phi^2 )^2
\ee
such that the Lagrangian becomes
\be
{\cal L} = \frac{1}{24}[8X^3 -5 (1-\phi^2)^6] +\frac{1}{4} (1-\phi^2)^2 [4X^2 +3(1-\phi^2)^4] +\frac{1}{2} [1+(1-\phi^2)^4][2X-(1-\phi^2)^2] .
\ee
The once integrated field equation for static solutions is equivalent to the condition that the one-one component of the energy momentum tensor is constant (see, e.g. \cite{bazeia3, bazeia2}). Further, for finite energy solutions this constant must be zero,
\be \label{T11-eq}
T_{11} = {\cal L}- 2X {\cal L}_X =0
\ee
where now $X=-\frac{1}{2}\phi'^2$ because $\phi$ is a static configuration. For the concrete example this equation reads
\be
-\frac{5}{3}X^3 -3(1-\phi^2)^2 X^2 - [1+(1-\phi^2)^4]X + \frac{1}{24}(1-\phi^2)^6 - \frac{1}{2}(1-\phi^2)^2 =0.
\ee
It may be checked easily that this equation is solved by $X=-\frac{1}{2}(1-\phi^2)^2$, i.e., $\phi '^2 = (1-\phi^2)^2$ which is just the field equation of the $\phi^4$ kink with the solution $\phi (x) = \pm\tanh (x-x_0)$. Therefore, our concrete example has the standard $\phi^4$ kink    
as a defect solution (it was specifically chosen to have this solution). Due to the nonlinear character of the above field equation, the model probably has more solutions (like the ones of Section 3), but this issue is beyond the scope of the present paper.   

Finally, let us remark that, although for the above model (i.e., the choice (\ref{alphas}) for the $\alpha_k$) the energy density is not positive semi-definite, it is easy to find a small variation of the model with positive semi-definite energy density. All one has to do is to increase the relative size of $\alpha_3$ and (or) $\alpha_1$ as compared to $\alpha_2$. Choosing, for example, $\alpha_1 $ and $\alpha_2$ as in (\ref{alphas}), and $\alpha_3 = \frac{1}{3}$, the resulting energy density is positive semi-definite for arbitrary $F(\phi)$, as may be shown easily.  If we further choose $F^2\sim (1-\phi^2)^2$ then the resulting potential will have at least the two vacua $\phi = \pm1$, and a linearization of the model near these two vacua shows that the vacua are approached exponentially, like in the case of the standard kink. Therefore, the model most likely supports kinks which interpolate between the two vacua, although the explicit kink solutions will be more complicated. 

\subsection{The models of Bazeia, Menezes and Petrov}

The supersymmetric K field theories of Bazeia, Menezes and Petrov (BMP) \cite{bazeia2} are based on the superfield
\be \label{chi}
\pa_\mu \Phi \pa^\mu \Phi = \pa_\mu \phi \pa^\mu \phi + 2 \theta^\alpha \pa_\mu \phi \pa^\mu \psi_\alpha -  2\theta^2 \pa_\mu \phi \pa^\mu F -\theta^2 \pa_\mu \psi^\alpha \pa^\mu \psi_\alpha .
\ee
Indeed, the bosonic component of the superfield $D_\alpha \Phi D^\alpha \Phi$ only consists of a term proportional to $\theta^2$, therefore multiplying this superfield by an arbitrary function of the above superfield (\ref{chi}), $f(\pa_\mu \Phi \pa^\mu \Phi)$, only the theta independent term $f(\pa_\mu \phi \pa^\mu \phi )$ will contribute, leading to the Lagrangian
\be \label{Baz}
{\cal L}_{\rm BMP} = -\left( D^2 [f(\pa_\mu \Phi \pa^\mu \Phi )\frac{1}{2} D_\alpha \Phi D^\alpha \Phi ] | \right) _{\psi =0} 
= f(\pa_\mu \phi \pa^\mu \phi )(F^2 + \pa_\mu \phi \pa^\mu \phi ).
\ee
Obviously, these Lagrangians produce a coupling of the auxiliary field $F$ with the kinetic term $\pa_\mu \phi \pa^\mu \phi$. On the other hand, the auxilliary field only appears quadratically, implying a linear (algebraic) field equation for $F$. 

First of all, we want to remark that for functions $f(\pa_\mu \phi \pa^\mu \phi)$ which have a Taylor expansion about zero, the same bosonic Lagrangians may be constructed from the building blocks (\ref{building-b}) of Section 2 by taking a different linear combination
(the fermionic parts of the corresponding Lagrangians will in general not coincide)
\begin{eqnarray}
{\cal L}^{(k)}_{\rm BMP} &\equiv & 
({\cal L}^{(k,0)})_{\psi =0}-\binom{k-1}{1}({\cal L}^{(k-1,1)})_{\psi =0}
+\binom{k-1}{2}({\cal L}^{(k-2,2)})_{\psi =0}+ \ldots \nonumber
\\
&& \ldots +(-1)^{k-1}\binom{k-1}{k-1}({\cal L}^{(1,k-1)})_{\psi =0} = 
(\partial^\mu\phi \partial_\mu\phi + F^{2}) 
(\pa_\mu \phi \pa^\mu \phi )^{k-1} . 
\label{k-n-sum2}
\end{eqnarray}
We may easily recover the Lagrangian (\ref{Baz}) by taking linear combinations of these,
\be
 {\cal L}_{\rm BMP} = \sum_{k=1}^\infty \beta_k {\cal L}^{(k)}_{\rm BMP} = (F^2 + \pa_\mu \phi \pa^\mu \phi ) \sum_k \beta_k (\pa_\mu \phi \pa^\mu \phi )^{k-1} \equiv (F^2 + \pa_\mu \phi \pa^\mu \phi ) f(\pa_\mu\phi \pa^\mu \phi) .
\ee
In order to have more interesting solutions, BMP added a potential term, as we did in Section 2. The resulting theories can, in fact, be analyzed with methods very similar to the ones employed in the previous sections. Concretely, they studied the Lagrangians
\be
{\cal L}_{\rm BMP}^{(P)} = f(\pa_\mu \phi \pa^\mu \phi ) (F^2 + \pa_\mu \phi \pa^\mu \phi ) + P' (\phi )F
\ee
which after eliminating the auxiliary field $F$ using its algebraic field equation
\be
F= -\frac{P'}{2f}
\ee
becomes
\be
{\cal L}_{\rm BMP}^{(P)} = f (\frac{P'^2}{4f^2} + \pa_\mu \phi \pa^\mu \phi ) -
\frac{P'^2 }{2f} = \pa_\mu \phi \pa^\mu \phi f - \frac{P'^2}{4f}
\ee
where we suppressed the arguments of $P$ and $f$ in the last expression to improve readability. The $X$ derivative of this Lagrangian is
\be
({\cal L}_{\rm BMP}^{(P)})_{,X} = f_{,X} \left( \frac{P'^2}{4f^2} + 2X\right) +2f
\ee
(please remember that $X\equiv \frac{1}{2}\pa_\mu \phi \pa^\mu \phi $ and $f=f(2X)$ such that $f_{,X} = 2f'$). The null energy condition already imposes rather nontrivial restrictions on the function $f$. A sufficient condition is $f\ge 0$, $f_{,X}\ge 0$ and $f \ge|Xf_{,X}|$ as may be checked easily. Finally, the once integrated field equation (\ref{T11-eq}) for static fields, after a simple calculation, leads to 
\be
\frac{1}{4f^2}(f+2Xf_{,X}) (8Xf^2 + P'^2)=0
\ee
where now $X=-\frac{1}{2}\phi'^2$ and $f=f(-\phi'^2)$ and, therefore, to the two equations
\be \label{BPS-eq}
2\phi ' (x)f(-\phi '(x)^2)= \pm P' (\phi)
\ee
where we reinserted the arguments in the last expression for the sake of clarity. For some choices of $f$ and $P$ these equations lead to defect solutions.
Finally, in the models of BMP the energy of a kink may be calculated with the help of the first order formalism first introduced in \cite{bazeia3}, in close analogy to the calculations presented in Section 3.A. It also remains true that, like in Section 3.A, the prepotential $P(\phi)$ is equal to the integrating function $W(\phi)$. The energy functional for static configurations (but {\em without} the use of the field equation) may, in fact, be re-written in a BPS form (exactly like for the standard supersymmetric scalar field theory), from which both the first order equations and the equality $P=W$ follow immediately. Indeed, the energy functional may be written like
\be
{ E}_{\rm BMP}^{(P)} = \int dx \left( \phi '^2 f + \frac{P'^2}{4f}\right) = \int dx \left( \frac{1}{4f}(2\phi ' f \mp P')^2 \pm \phi ' P' \right)
\ee
and for a solution to the first order (or BPS) equation (\ref{BPS-eq}) (we take the plus sign for definiteness) the resulting energy is therefore 
\be
{ E}_{\rm BMP}^{(P)} = \int_{-\infty}^\infty dx \phi ' P' = \int_{\phi (-\infty)}^{\phi (\infty)} d\phi P' =P(\phi (\infty)) - P(\phi (-\infty))
\ee
which proves the above statement. 
 For a more detailed discussion we refer to \cite{bazeia2} (we remark that BMP use a slightly different notation in \cite{bazeia2}: they use the notation $h$ instead of $P$ for the prepotential, and their $X$ is defined like $X=\pa_\mu \phi \pa^\mu \phi$ instead of the definition $X=\frac{1}{2} \pa_\mu\phi \pa^\mu \phi $ used in the present paper and in \cite{bazeia3}).

\section{Discussion}

We developed and described a method to construct general supersymmetric scalar K field theories in 1+1 and 2+1 dimensions. Among these theories, we found a large class of models which, in the purely bosonic sector, consist of a generalized kinetic term plus a potential, where the vacuum structure of the potential is crucial for the determination of the topological defect solutions, similarly to the standard case (i.e., kinetic term $\sim \pa_\mu \phi \pa^\mu \phi$). Due to the enhanced nonlinearity of these supersymmetric K field models there are, nevertheless, some significant differences, like different roots of the first order field equations leading to a larger number of kink solutions, or the possibility to join different solutions forming additional kinks in the space ${\cal C}^1$ of continuous functions with a continuous first 
derivative \footnote{  
Whether such ${\cal C}^1$ solutions are physically relevant depends, of course, on the concrete physical system under consideration, but we want to emphasize that if the supersymmetric scalar field theory is interpreted as an {\em effective} theory then the ${\cal C}^1$ solutions should be taken into account. In this case, the spike in the field derivative (and in the energy density) of a ${\cal C}^1$ solution will, in any case, be resolved by the true UV degrees of freedom.}. 
These results are new and are by themselves interesting, broadening the range of applicability of supersymmetry to a new class of field theories and, at the same time, enhancing our understanding of these field theories. 

As far as possible applications are concerned, the natural arena seems to be the area of cosmology, as already briefly mentioned in the introduction. Indeed, if these scalar field theories are interpreted as effective theories which derive from a more fundamental theory with supersymmetry (like string theory), then it is natural to study the supersymmetric versions of the effective models. If, in addition, the defect formation and phase transition (e.g. from a symmetry breaking phase to a symmetric phase) relevant for cosmological considerations occur at time or energy scales where supersymmetry is still assumed unbroken, then also the defect solutions of the {\em supersymmetric} effective field theories are the relevant ones.      

At this point, several questions show up. The first one is the inclusion of fermions. It is, e.g., expected that, as a consequence of supersymmetry and translational invariance, there should exist a fermionic zero mode for each kink background where, in addition, the fermionic zero mode is equal to the spatial derivative of the kink. This fact has already been confirmed explicitly in some supersymmetric K field theories \cite{bazeia2}, \cite{susy1}. The inclusion of fermions in the Lagrangians studied in the present article does not present any difficulty on a fundamental level, the only practical obstacle being that, for purely combinatorical reasons, the resulting expressions will be rather lengthy. A second question is whether the SUSY algebra in a kink background contains central extensions related to the topological charge of the kink, as happens for the standard supersymmetric kink \cite{witten-olive}. Again, this problem has already been studied for some supersymmetric K field theories \cite{susy1}.
A further question concerns the issue of quantization. If the SUSY K field theories are interpreted as {\em effective} theories, as would be appropiate, e.g., in a cosmological context, then already the classical model contains some relevant information of the underlying quantum theory, like spontaneous symmetry breaking or the existence of topological defects.  In this context, the quantization of quadratic fluctuations about the topological defect is the correct procedure to obtain further information about the underlying theory. Independently, one may, nevertheless, pose the problem of a full quantization of the supersymmetric K field theory, where the enhanced degreee of nonlinearity certainly implies further complications. One may ask, e.g.,. whether the additional, non-quadratic kinetic terms may be treated perturbatively, like the non-quadratic terms of the potential in the standard case. The answer will certainly depend on the space-time dimension. A related question is whether supersymmetry simplifies the task by taming possible divergences, as happens in the standard case. Here it is interesting to note that, even at the classical level, supersymmetry implies some restrictions on possible Lagrangians which are visible already in the bosonic sector. Indeed, as is obvious e.g. from Eq. (\ref{L-b-F}), there exists a relation between the kinetic and the potential terms (this relation is responsible for the existence of the so-called generic solutions). One wonders what this relation implies for the quantum theory, e.g., in the form of Ward-like identities.
These and related questions will be investigated in future publications.

Finally, we think that the supersymmetric models we found present some independent mathematical interest of their own, given their high degree of non-linearity, on the one hand, and the possibility to obtain rather precise information on their solutions (e.g. all kink solutions together with their exact energies), on the other hand. Further investigations in this direction (e.g., time-dependent solutions, or the stability of topological solitons) will be pursued, as well.    

\section*{Acknowledgement}
The authors acknowledge financial support from the Ministry of Science and Investigation, Spain (grant FPA2008-01177), 
the Xunta de Galicia (grant INCITE09.296.035PR and
Conselleria de Educacion), the
Spanish Consolider-Ingenio 2010 Programme CPAN (CSD2007-00042), and FEDER.

\end{document}